# Characterization of cutting-edge CMOS Active Pixel sensors within the CYGNO Experiment

B.D. Almeida[1], F.D. Amaro [2], R. Antonietti [3,4], E. Baracchini [5,6], L. Benussi [7], S. Bianco [7], C. Capoccia [7], M. Caponero [7,8], L.G.M de Carvalho [1], G. Cavoto [9,10], I.A. Costa [7], A. Croce[7], M. D'Astolfo [5,6], G. D'Imperio [10], E. Danè [7], G. Dho [7], E. Di Marco [10], J.M.F. dos Santos [2], D. Fiorina [5,6], F. Iacoangeli [10], Z. Islam [5,6], E. Kemp [11], H. P. Lima Jr [5,6], G. Maccarrone [7], R.D.P. Mano [2], D. J. G. Marques [5,6], G. Mazzitelli [7], P. Meloni [3,4], A. Messina [9,10], C.M.B. Monteiro [2], R.A. Nobrega*[1], I.F. Pains [1], E. Paoletti[7], F. Petrucci [3,4], S. Piacentini [5,6], D. Pierluigi[7], D. Pinci [10], F. Renga [10], R.J.C. Roque [2], A. Russo[7], G. Saviano[7,12], P.A.O.C. Silva [2], N.J. Spooner[13], R. Tesauro [7], S. Tomassini [7], and D. Tozzi [9,10]

[1] *Universidade Federal de Juiz de Fora; Faculdade de Engenharia; 36036-900; Juiz de Fora; MG; Brazil*
[2] *LIBPhys; Department of Physics; University of Coimbra; 3004-516 Coimbra; Portugal*
[3] *Dipartimento di Matematica e Fisica; Università Roma Tre; 00146; Roma; Italy*
[4] *Istituto Nazionale di Fisica Nucleare; Sezione di Roma Tre; 00146; Rome; Italy*
[5] *Gran Sasso Science Institute; 67100; L'Aquila; Italy*
[6] *Istituto Nazionale di Fisica Nucleare; Laboratori Nazionali del Gran Sasso; 67100; Assergi; Italy*
[7] *Istituto Nazionale di Fisica Nucleare; Laboratori Nazionali di Frascati; 00044; Frascati; Italy*
[8] *ENEA Centro Ricerche Frascati; 00044; Frascati; Italy*
[9] *Dipartimento di Fisica; Sapienza Università di Roma; 00185; Roma; Italy*
[10] *Istituto Nazionale di Fisica Nucleare; Sezione di Roma; 00185; Roma; Italy*
[11] *Universidade Estadual de Campinas - UNICAMP; Campinas 13083-859; SP; Brazil*
[12] *Dipartimento di Ingegneria Chimica; Materiali e Ambiente; Sapienza Università di Roma; 00185; Roma; Italy*
[13] *Department of Physics and Astronomy; University of Sheffield; Sheffield; S3 7RH; UK*

**Abstract**

Time Projection Chambers equipped with Gas Electron Multipliers and optical readout by scientific CMOS cameras are a promising technology for low-energy particle detection, as demonstrated by the CYGNO experiment. As a step toward identifying the optimal CYGNO detector configuration, we performed a detailed characterization of two state-of-the-art scientific CMOS sensors focusing on dark signal behavior across a range of exposure times and on detection sensitivity, assessed using the well-defined X-ray emissions from a $^{55}Fe$ radioactive source, which reproduce the low-light conditions expected in the CYGNO experiment. CYGNO currently employs a very low-noise sensor produced by Hamamatsu, the ORCA-Fusion, for testing and validation of its detection system. However, Hamamatsu has recently introduced two new sensors that could be of interest for future upgrades. The first is an upgraded model of the current sensor, called ORCA-Fusion-BT. It features a back-illuminated design, enabling it to reach a quantum efficiency of up to 95% at 550 nm. The second is a next-generation scientific sensor referred to as the ORCA-Quest. While its peak quantum efficiency is not as high as that of the ORCA-Fusion-BT, it offers high sensitivity across a broader spectral range, extending into the ultraviolet region, which may be beneficial for many scientific applications. In addition, it delivers ultra-low readout noise of just 0.27 electrons, which is approximately 2.6 times lower than that of the Fusion family. These two sensors therefore represent a significant opportunity to enhance the performance of scientific experiments, including those conducted by the CYGNO collaboration. In this scenario, this document presents a comprehensive characterization of these new sensors to evaluate their potential relevance for scientific experiments requiring high performance in photon-limited environments, as well as to assess their suitability for integration into the CYGNO detector system.

*Corresponding author: rafael.nobrega@ufjf.br

# 1 Introduction

## 1.1 Scientific context and motivation

Advances in low-light imaging technologies are increasingly essential for particle and astroparticle physics [1, 2], where precise detection of faint optical signals directly determines experimental sensitivity and resolution. The CYGNO experiment [3] is pursuing the development of an optically readout gaseous Time Projection Chamber (TPC) for rare-event searches in astroparticle physics, such as Dark Matter direct detection. Its detection principle is based on a $He/CF_4$ gas mixture coupled to a triple-GEM amplification stage, where the ionization electrons produced in the drift region undergo avalanche multiplication, leading to light emission. This light is recorded by high-resolution scientific CMOS (sCMOS) cameras equipped with fast optics, as well as by photomultiplier tubes (PMTs), enabling precise three-dimensional reconstruction of particle tracks. Since the imaging sensors are the core readout element, their performance directly impacts the detector's sensitivity, track reconstruction fidelity, and background rejection capability. In the context of dark matter detection, precise track reconstruction is essential, as it provides not only the energy deposition but also the orientation of nuclear recoils. Directional sensitivity represents one of the most powerful tools for identifying dark matter candidates, since the distribution of nuclear recoils induced by potential dark matter particles is expected to exhibit a characteristic anisotropy aligned with the motion of the Solar System through the Galactic halo. Achieving low-noise operation and accurate imaging is therefore essential to maximize the detector's discrimination capability and discovery potential.

Manufacturer specifications provide important baseline parameters, but these are typically obtained under idealized laboratory conditions and may not accurately reflect the performance within real detector configurations. To address this gap, the present work provides a comprehensive characterization of state-of-the-art image sensors. In addition to standard performance metrics, measurements not typically reported by manufacturers are presented, including the spatial distribution of noise and pedestal levels across a wide range of exposure times, as well as tests performed with radioactive sources to mimic realistic experimental conditions. These measurements extend beyond controlled setups to capture sensor behavior under operational constraints relevant to gaseous TPC detectors and other photon-limited systems. Furthermore, direct experiment-specific benchmarking enables a quantitative comparison of candidate devices, offering practical guidance on which sensors deliver superior performance when integrated into a scientific apparatus. While the primary goal is to evaluate the suitability of these sensors for integration into the CYGNO detector system, the results have broader relevance for a wide range of photon-limited scientific applications.

In summary, achieving the scientific goals of the CYGNO collaboration requires advanced imaging sensors capable of capturing particle tracks with high spatial detail and sensitivity. Scientific CMOS technology offers several advantages that make it particularly well suited to the demanding detection requirements of the experiment. The main characteristics that an imaging sensor must possess to operate effectively in this context are outlined below:

- **Low-light sensitivity**: In the detector environment, ionization events typically result in very weak light emission, especially in low-energy interactions. Therefore, it is desirable that the sCMOS sensor be sensitive under low-light conditions, capturing even the faintest signals. This level of sensitivity is ensured by the high quantum efficiency of the sensor, which should be greatest in the wavelength range relevant to the experiment.

- **Low readout noise**: Minimizing readout noise is a crucial requirement for any sensor employed in low-light experiments. Electronic noise can mask or distort the weak signals generated by ionization events, compromising data accuracy. State-of-the-art sCMOS sensors feature extremely low readout noise, ensuring an excellent signal-to-noise ratio.

- **High spatial resolution**: For the CYGNO experiment, it is essential that the sCMOS sensor provides high spatial resolution. This characteristic allows precise capture of the tracks of ionizing particles passing through the gas within the detection chamber. A sensor capable of distinguishing fine details is crucial for identifying individual tracks and precisely characterizing their properties, including direction, orientation, energy, and length.

Selecting a suitable sensor for the CYGNO detector requires considering the combination of these technical features, as they are fundamental for obtaining reliable and detailed data in the search for dark matter signals and other rare phenomena.

## 1.2 Key properties of the sensors under investigation

The latest CYGNO detector prototype, named LIME (Long Imaging Module Experiment) [4, 5], has been employing the ORCA-Fusion (C14440-20UP) [6, 7] sensor for system characterization and validation. However, state-of-the-art advances in scientific CMOS technology have led Hamamatsu to develop two new sensors: ORCA-Fusion-BT (C15440-20UP) [6] and ORCA-Quest (C15550-20UP) [8], hereafter referred to simply as Fusion-BT and Quest.

Fusion-BT features a back-illuminated architecture, enabling quantum efficiency improvements over previous generations and attaining values up to 95% at 550 nm. This sensor was developed to deliver superior performance under low-light conditions, making it particularly suitable for areas such as fluorescence microscopy, live-cell imaging, and other scientific applications that demand high precision and data reliability. With a resolution of 5.3 megapixels, consisting of a 2304 × 2304 pixel array, the ORCA-Fusion-BT sensor has pixels measuring 6.5 $\mu$m × 6.5 $\mu$m, allowing the capture of high-definition images with excellent detail fidelity. The ORCA-Fusion-BT is characterized by its low readout noise, which can be as low as 0.7 electrons RMS. In addition to low noise, the ORCA-Fusion-BT sensor can operate at a frame rate of up to 89.1 FPS at full resolution, making it ideal for capturing fast and dynamic events without sacrificing image resolution or quality.

The ORCA-Quest stands out for its extremely low readout noise, specified at 0.27 electrons RMS, which is approximately 2.6 times lower than the Fusion family. With a resolution of 9.4 megapixels, distributed across a 4096 × 2304 pixel array, it offers high pixel density, with each pixel measuring 4.6 $\mu$m × 4.6 $\mu$m. This configuration allows capturing high-definition images with precise details, which is essential for applications such as super-resolution microscopy and materials analysis. The back-illuminated architecture of the sensor allows for increased quantum efficiency, reaching up to 85% at 460 nm, ensuring excellent light sensitivity across a wide wavelength range, including the near-infrared region. Quest is equipped with an advanced real-time pixel correction function, ensuring exceptional uniformity across the pixel array. This feature is essential for applications requiring high precision and data consistency, as it eliminates unwanted pixel-to-pixel variations, resulting in more uniform images and more reliable data. Another important technical aspect is the sensor's ability to resolve the number of photoelectrons generated in each pixel. Thanks to the low readout noise and high uniformity, it can accurately count photoelectrons, which is critical for applications where light quantification is essential, such as spectroscopy and quantum imaging.

Table 1 shows the main characteristics of Fusion, Fusion-BT and Quest sensors.

Table 1: Comparison between Fusion, Fusion-BT and Quest sensors.

| **Feature** | **ORCA-Fusion** | **ORCA-Fusion-BT** | **ORCA-Quest** |
| --- | --- | --- | --- |
| Resolution | 2304 × 2304 (5.3 MP) | 2304 × 2304 (5.3 MP) | 4096 × 2304 (9.4 MP) |
| Pixel size | 6.5 $\mu$m × 6.5 $\mu$m | 6.5 $\mu$m × 6.5 $\mu$m | 4.6 $\mu$m × 4.6 $\mu$m |
| Quantum Efficiency | up to 80% @ 550 nm | up to 95% @ 550 nm | up to 85% @ 460 nm |
| Readout noise (RMS) | 0.7 e$^-$ (Ultra Quiet Scan) | 0.7 e$^-$ (Ultra Quiet Scan) | 0.27 e$^-$ (Ultra Quiet Scan) |
| Dark current | 0.2 e$^-$/pixel/second | 0.3 e$^-$/pixel/second | 0.016 e$^-$/pixel/second |
| Max frame rate | up to 89 fps | up to 89 fps | up to 30 fps |

As noted previously, manufacturers' specifications provide useful baseline information, but supplementary measurements are required for a more complete understanding of sensor performance. In particular, dark signal measurements allow for a detailed assessment of baseline offsets and noise levels as a function of exposure time. Complementary measurements performed with radioactive sources enable the study of the sensor response to low-light events representative of those expected in the CYGNO experiment. Together, these measurements provide a comprehensive verification of key parameters and their impact on detector performance, ensuring that the device operates as required in the experimental context. Moreover, the methodology adopted here may also be relevant to other scientific experiments that rely on high-performance imaging sensors for detecting faint light signals.

The structure of this paper is as follows. Section 2 describes the measurement setup, including the employed detector and its main characteristics. The experimental results are then presented in Sections 3 and 4. Section 3 reports the characterization of the dark signal as a function of exposure time, providing a detailed analysis of its behavior through statistical and spatial analyses to provide both a general and spatial understanding of the sensor-inherent background signal. Section 4 examines the impact of the sensor characteristics on the CYGNO experiment, with emphasis on light-related measurements performed using a $^{55}Fe$ radioactive source. In this section, in addition to evaluating the CYGNO detector equipped with the Fusion-BT and Quest sensors, we also compare their performance with that obtained using the Fusion sensor. Finally, Section 5 offers concluding remarks and summarizes the main findings of this work.

## 2 Measurement setup

While the dark signal analysis is performed by ensuring that the sensors are not exposed to any light, evaluating their impact on CYGNO requires the use of the experiment's detection apparatus. In this section, we describe the employed setup, along with the key characteristics relevant to the proposed sensor-related measurements.

## 2.1 LIME detector

A schematic of the LIME detector [4] is shown in Figure 1. It is the largest TPC developed to date within the CYGNO project. Built at the National Laboratory of Frascati (LNF), it features a sensitive volume of 50 liters and a drift length of 50 cm. The gas container of LIME is constructed from *plexiglass* and is filled with a scintillating $He/CF_4$ mixture in a 60/40 ratio at room temperature and atmospheric pressure. At the top, an ETFE (*ethylene-tetrafluoroethylene*) window was installed, with a thickness of 125 $\mu$m and a width of 5 cm, extending along the full length of the field cage. This window was specifically designed to allow tests with low-energy radioactive sources, taking advantage of ETFE's mechanical and thermal properties.

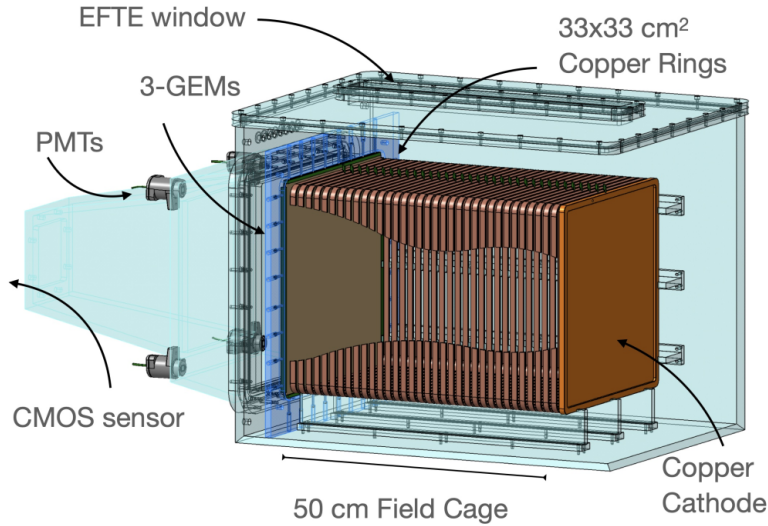

Figure 1: Concept drawing of the LIME prototype.

The LIME readout system consists of a triple-GEM structure with a surface area of 33 × 33 cm$^2$. Each GEM layer is biased at a nominal voltage of 440 V, creating an electric field sufficient to induce electron multiplication within the GEM holes. The resulting avalanche process amplifies the signal produced by particle interactions in the gas, while the triple-GEM configuration ensures both high efficiency and stable gain. During the electron multiplication process, light is produced and subsequently collected by the imaging sensor and the four photomultiplier tubes. On the opposite side of the Triple-GEM system is the detector cathode, responsible for collecting the positive ions generated by the interaction of gas molecules with incident particles. The TPC drift volume is defined by a field cage positioned between the cathode, made of a 3 mm thick copper layer, and the GEM-stack anode. A cathode–anode potential difference of 50 kV is applied, with the field cage consisting of copper rings arranged to establish a uniform electric field throughout the detector's sensitive volume. The electric field established between the cathode and the nearest GEM, known as the drift field, directs the electrons released by particle interactions in the gas toward the readout system.

## 2.2 LIME light collectors

Nominally, LIME employs an sCMOS imaging sensor manufactured by Hamamatsu, model ORCA-Fusion, positioned 62 cm away from the GEM surface. For optimal light collection, the sensor is coupled to Schneider-Kreuznach lenses [9], featuring a focal length of 25.6 mm and a maximum aperture ratio of $f_\# = 0.95$.

In addition to the sCMOS camera, LIME uses four photomultiplier tubes symmetrically positioned around the sCMOS sensor at a distance of approximately 25 cm from the GEM surface. The PMTs, model Hamamatsu R7378A [10], operate at a maximum voltage of 1200 V and have a gain of $2.0 \times 10^6$, being responsible for detecting photons generated during the electron avalanche process in the GEMs. The combination of PMTs and sCMOS allows three-dimensional reconstruction of detected events, providing detailed information on the spatial and temporal extent of the signals.

## 2.3 Fraction of the photons received by the imaging sensor

The sensor-to-GEM distance is determined by the sensor dimensions and its placement within LIME to ensure full coverage of the $33 \times 33$ cm$^2$ GEM plane. While the Fusion family sensors feature a square active area of $14.976 \times 14.976$ mm$^2$, the largest square active area available for the Quest sensor is $10.598 \times 10.598$ mm$^2$.

The fraction of photons received by each sensor can be estimated using a thin-lens approximation. By simplifying the lens as a single converging lens, Equation 1 can be used, where $\epsilon$ represents the fraction of photons received by the sensor from an isotropic point light source, $L$ corresponds to the lateral size of the GEM plane, and $\ell$ to that of the sensor, respectively. $f_\#$ is the optical system's *f-number*, defined in Section 2.2. The sensor covers an area of $35 \times 35$ cm$^2$, corresponding to the GEM plane size plus a 1 cm margin on each side.

$$\epsilon = \frac{1}{[4f_\# \left(\frac{L}{\ell} + 1\right)]^2} \quad (1)$$

Table 2 lists the sensor side lengths, their distance from the GEM plane, and the corresponding fraction of collected photons, considering the sensors of interest for the CYGNO collaboration.

| **Sensor** | **Side length** (cm) | **GEM distance** (cm) | **Fraction of photons** |
|---|---|---|---|
| Fusion | 1.50 | 62 | $1.17 \times 10^{-4}$ |
| Fusion-BT | 1.50 | 62 | $1.17 \times 10^{-4}$ |
| Quest | 1.06 | 87 | $5.98 \times 10^{-5}$ |

Table 2: Side length, distance from the GEM plane, and photon fraction for each of the tested sensors.

As observed, the Fusion and Fusion-BT sensors can be positioned closer to the GEM plane, resulting in approximately 1.95 times more photons reaching the sensor compared to the Quest sensor.

## 2.4 Sensitivity to photons emitted toward the sensor

Regarding the sensitivity to photons propagating toward the imaging sensors, some key components must be considered: the photon energy spectrum emitted by the GEM, the transmittance of the detector's glass (*Plexiglass*) layer between the GEM plane and the sensor [11], and the quantum efficiency of the considered sensor. Figure 2 shows these components, and Table 3 presents the percentages of photons emitted toward the sensor that are converted into photoelectrons—considering only quantum efficiency in the middle column, and quantum efficiency combined with the Plexiglass effect in the last column.

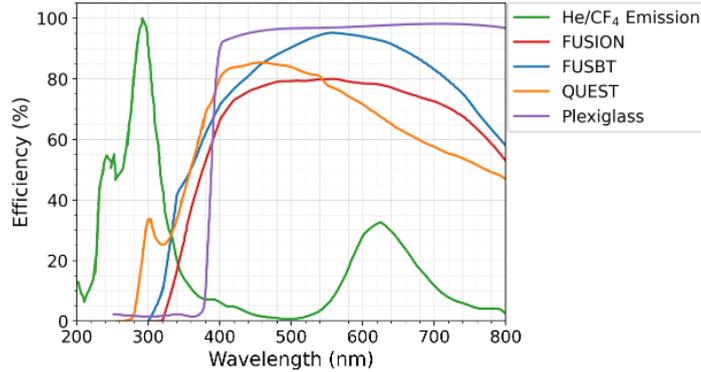

Figure 2: Photon energy spectrum emitted by the GEM (green), transmittance of the *Plexiglass* (violet), and the Quantum Efficiency of each sensor (red, blue, and orange).

| Sensor | Quantum Efficiency | Quantum Efficiency + Plexiglass |
|---|---|---|
| Fusion | 29.51% | 26.89% |
| Fusion-BT | 36.14% | 31.77% |
| Quest | 34.47% | 23.97% |

Table 3: Percentage of photons emitted toward the sensor that are converted into photoelectrons, considering only Quantum Efficiency and Quantum Efficiency together with the *Plexiglass* effect.

According to Table 3, the Fusion-BT exhibits a photoelectron generation factor 1.32 times higher than that of the Quest sensor, when considering the values reported in the last column of the table. It is also worth noting that, without the *Plexiglass*, this ratio would decrease to 1.05, as the Quest sensor exhibits higher efficiency in the ultraviolet region, almost nullifying this factor related to photon-to-photoelectron conversion.

# 3 Sensor dark signal characteristics

The dark signal and its dependence on exposure time play a crucial role in the quality of acquired images, as increased dark signal can mask weak events of interest and degrade performance in low-light conditions. It can be characterized through pedestal measurements, which capture the baseline offset in the absence of light, and noise measurements, which describe fluctuations around this baseline. In low-light environments (or low event-rate conditions), longer exposure times are often used to allow a greater amount of light (or events) to be collected by the sensor, thereby increasing the number of captured photoelectrons. However, this increase in exposure time can also lead to higher noise levels, particularly due to the intrinsic thermal variation of the sensor during prolonged data collection periods.

In order to investigate in detail the relationship between noise and exposure time for the studied sensors, Fusion-BT and Quest, an experiment was conducted in dark conditions, in which images were acquired using seven distinct exposure times: 10, 30, 100, 300, 1000, 3000 and 10000 ms. For each exposure time, 400 images were acquired for each sensor under analysis, resulting in a total of 5600 images. This substantial dataset enables statistical analysis to be performed, providing a deep understanding of the dark signal characteristics of the sensors and the impact of exposure time.

The results in this section are divided into two parts: Section 3.1 presents the pedestal (baseline offset) measurements, while Section 3.2 provides a detailed description of the noise behavior of the sensors.

## 3.1 Pedestal analysis

Figure 3 shows the normalized distributions of the pedestal measurements for the Fusion-BT and Quest sensors for different exposure times. As expected, the Fusion-BT and Quest distributions are centered around 100 and 200 ADC counts, respectively, in agreement with the specifications provided in their datasheets. Both exhibit a right-hand tail that becomes more pronounced for longer exposure times. In particular, for the Quest sensor, the distribution remains practically unchanged up to 300 ms. The Fusion-BT sensor also exhibits a left-hand tail, although it is much less pronounced than the right-hand one.

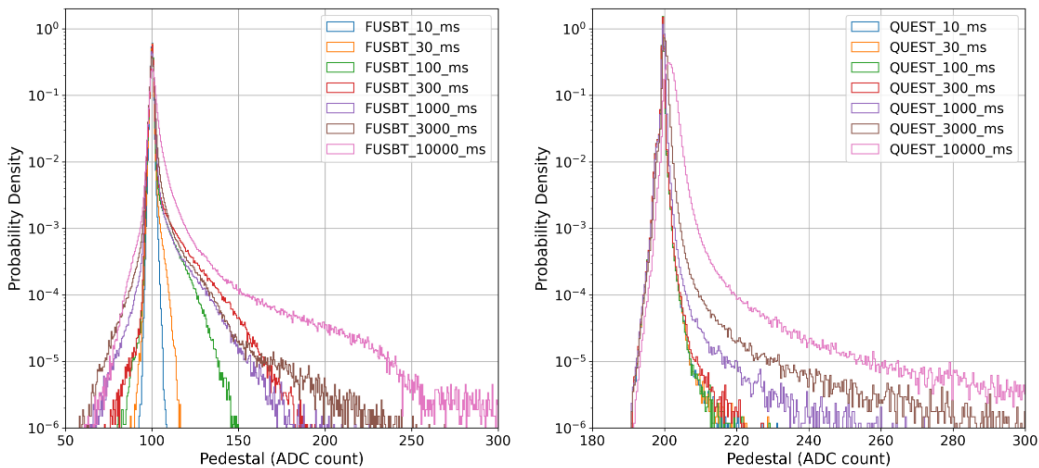

Figure 3: Pedestal distributions for different exposure times for the Fusion-BT and Quest sensors.

Figure 4 shows the evolution of the mean and standard deviation of the pedestal distributions for both sensors as a function of exposure time. To improve visualization, 100 and 200 ADC counts were subtracted from the mean measurements of Fusion-BT and Quest, respectively. It is clear from comparison with the distributions in Figure 3 that the increase in the measured pedestal mean and standard deviation is associated with the tails of the distributions, which broaden as the exposure time increases.

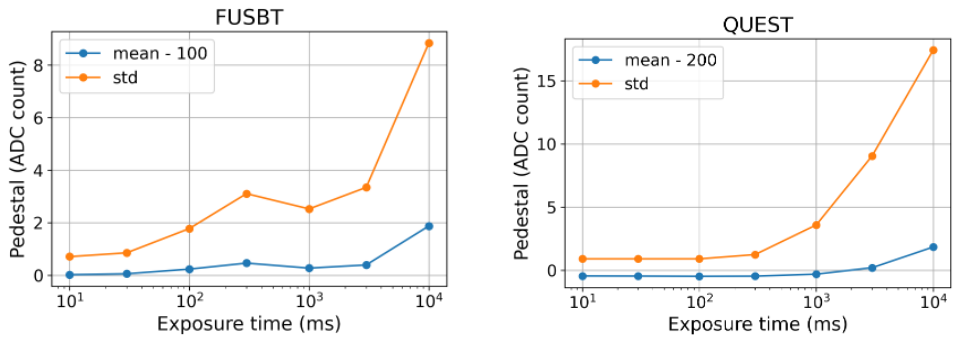

Figure 4: Mean and standard deviation of the pedestal as functions of exposure time for the Fusion-BT and Quest sensors.

Figure 5 shows pixel-wise pedestal measurements for the Fusion-BT sensor for exposure times of 100 ms, 1 000 ms, and 3 000 ms. The results clearly indicate that the increase in the pedestal distribution tails is primarily associated with the pixels located along the right border of the sensor's active area and in the lower-right corner, whose pedestal values become more pronounced as the exposure time increases. Additionally, a row fixed-pattern is noticeable, arising from slight variations in the readout electronics of each row— a characteristic feature of sCMOS image sensors.

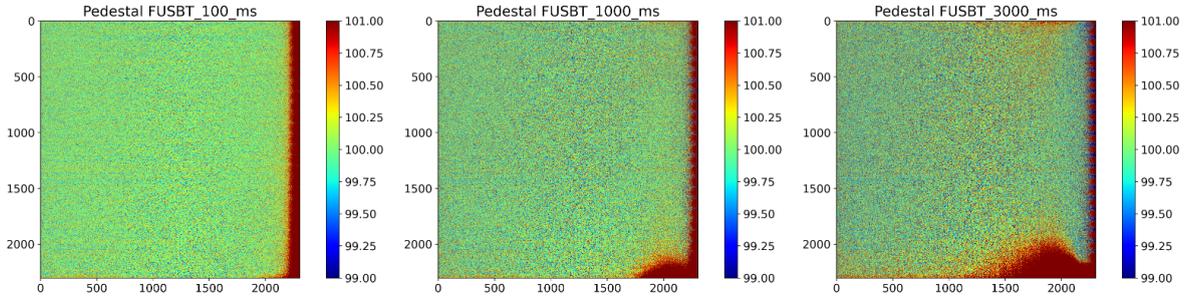

Figure 5: Pixel-wise pedestal measurements for the Fusion-BT sensor for exposure times of 100 ms, 1 000 ms, and 3 000 ms.

Figure 6 shows pixel-wise pedestal measurements of the Quest sensor for exposure times of 100 ms and 3000 ms. At the bottom, three specific 64×72-pixel regions for 3000 ms are displayed, selected from the upper-left, upper-central, and upper-right portions of the sensor. The results show no clearly defined region responsible for the increase in pedestal mean, except for the first pixels in the upper part of the active area, which exhibit a noticeable rise in pedestal values. Additionally, a column fixed-pattern is also apparent.

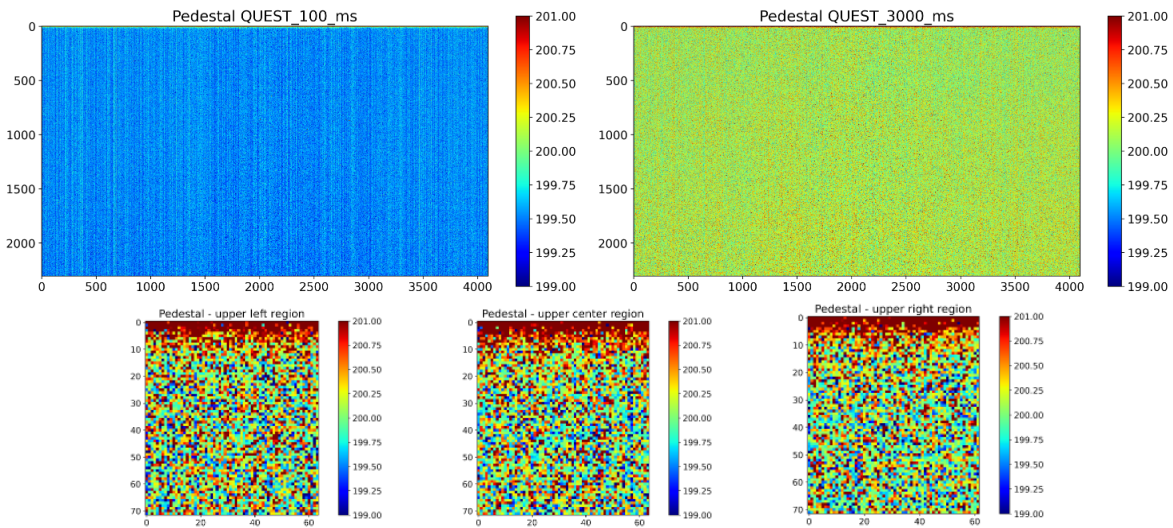

Figure 6: Pixel-wise pedestal measurements of the Quest sensor for exposure times of 100 ms and 3000 ms. At the bottom, three specific 64×72-pixel regions are shown, selected from the upper-left, upper-central, and upper-right regions of the sensor.

## 3.2 RMS noise analysis

Figure 7 shows the RMS noise distributions for different exposure times for the Fusion-BT and Quest sensors. As observed, for Fusion-BT, the right-hand tail becomes more pronounced as the exposure time increases. For the Quest sensor, the right-hand tail remains largely unchanged up to 1000 ms.

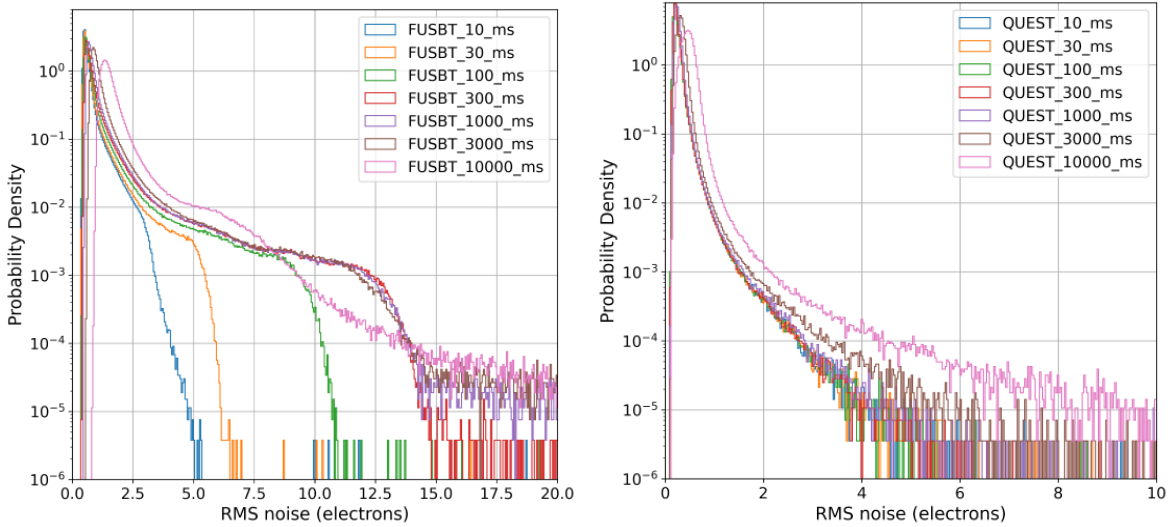

Figure 7: RMS noise distributions for different exposure times for the Fusion-BT and Quest sensors.

Figure 8 shows the RMS noise mean and standard deviation as a function of exposure time for both sensors. A comparison between the distributions in Figure 7 makes it clear that the observed increase in the RMS noise mean and standard deviation are associated with the right-hand tail of the distribution.

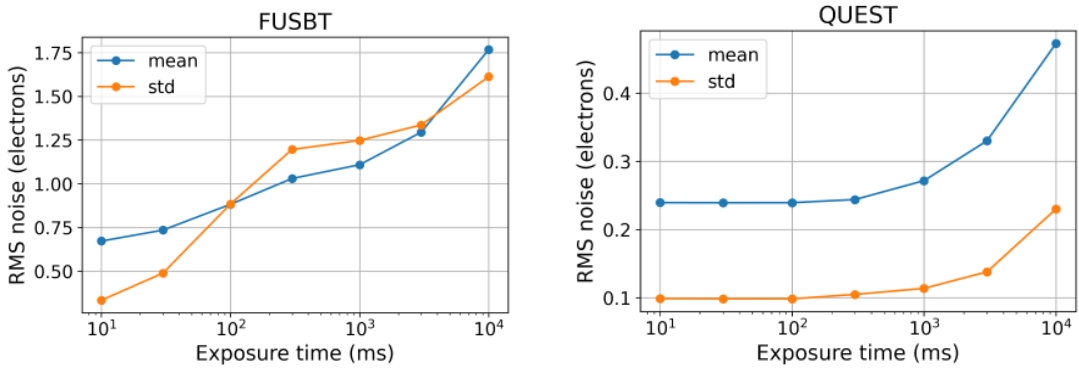

Figure 8: RMS noise mean and standard deviation as a function of exposure time for the Fusion-BT and Quest sensors.

Figure 9 shows the pixel-wise RMS noise measurements for exposure times of 100 ms, 1000 ms, and 3000 ms for the Fusion-BT sensor. As observed, the noise values vary more noticeably at the sensor edges than in the central regions. This variation is particularly evident along the right edge and in the lower-right corner of the sensor, where the noise values increase consistently with longer exposure times. Therefore, it can be concluded that the growth of the right-hand tails of the distributions shown on the left side of Figure 7 is associated with such observed edge effect.

For the Quest sensor, as shown by Figure 10, the noise pattern becomes much milder, being slightly noisier at the top for the 10 ms exposure time and slightly noisier at the bottom for the 10000 ms exposure time. As the exposure time increases, the color of the entire image changes in accordance with the RMS noise mean values shown in Figure 8. It therefore shows a considerable uniformity of RMS noise throughout the sensor's active area.

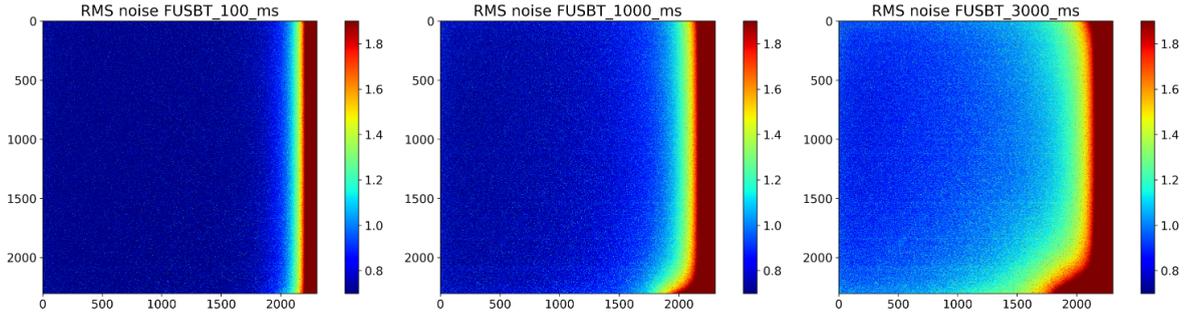

Figure 9: Pixel-wise RMS noise measurements for the Fusion-BT sensor for exposure times of 100 ms, 1000 ms, and 3000 ms.

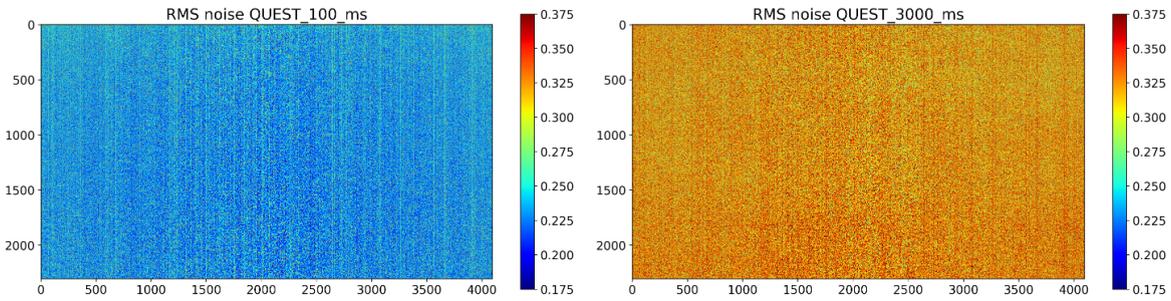

Figure 10: Pixel-wise RMS noise measurements for the Quest sensor for exposure times of 100 ms and 3000 ms.

To quantitatively illustrate how the RMS noise varies from the center toward the edges, the active area was divided into eight radial sectors. The first sector extends from the center to the right edge, the second from the center to the upper-right corner, and so on, covering the directions: right, upper-right, up, upper-left, left, lower-left, down, and lower-right. Each sector is composed of blocks measuring 64×64 pixels for the Fusion-BT sensor and 64×72 pixels for the Quest sensor. The blocks are arranged sequentially from the center outward in fixed steps in a way to form 32 blocks per sector, as illustrated in Figure 11.

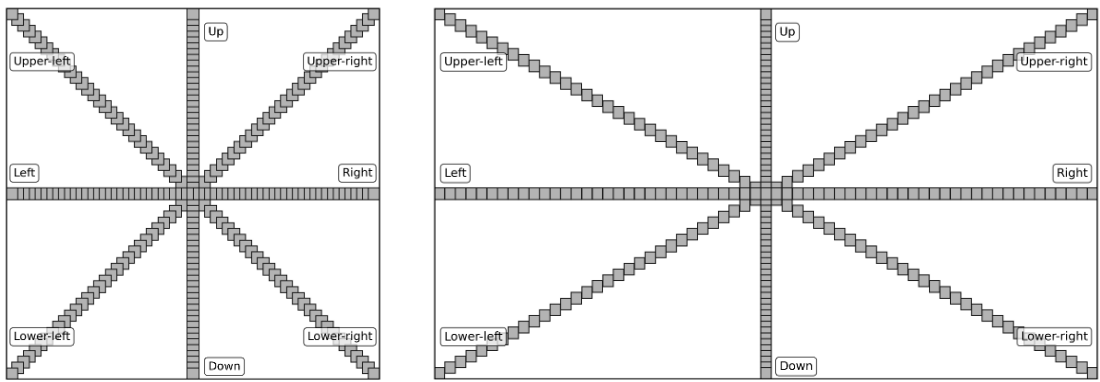

Figure 11: Visualization of the regions created from the center to the edge of the Fusion-BT and Quest sensors.

The variation of RMS noise across the eight sectors of the Fusion-BT sensor is shown in Figure 12. As observed, all sectors that cover the right side of the sensor's active area (right, upper-right and lower-right) show a pronounced increase in RMS noise near the border. The noise goes from 0.6 to almost 2 electrons for 10 ms of exposure time and becomes more pronounced for longer exposures, rising from 1.5 to approximately 7 electrons for 10000 ms. In particular, for the left sector, the RMS noise decreases gradually from the center toward the border of the sensor's active area, especially at higher exposure times. A similar trend is observed for the upper-left sector. In both cases, a slight rise in noise appears in the outermost blocks, suggesting a small increase in RMS noise at the left-side edge of the sensor's active area. Finally, when the exposure time is increased from 10 ms to 10000 ms across the seven defined steps, the RMS noise at the center of the sensor's active area correspondingly rises as follows: 0.62, 0.63, 0.66, 0.71, 0.78, 0.97, and 1.47 electrons.

Figure 13 presents the corresponding results for the Quest sensor. As can be seen, the border effect is practically absent. The RMS noise increases with exposure time, with the first significant change occurring between 300 ms and 1000 ms.

To better understand the variation of RMS noise throughout the sensor, a more detailed analysis is required. Figure 14 shows the RMS noise across the sensor's active area, which has been rebinned to form a 64x32 matrix. As can be seen, the RMS noise varies by less than 5% when moving from the center toward the borders. This analysis shows that for exposure times below 1000 ms, the highest RMS noise regions are located at the upper border of the sensor, whereas for exposure times above 1000 ms, they appear at the bottom. For 1000 ms, the entire border exhibits similar noise levels, slightly higher on the corners. The colorbar values in each image indicate an increase in RMS noise from one exposure time to the next. The range of the colorbar was chosen so that 5% of the values fall below the lower limit and 5% above the upper limit.

# 4 Impact on the CYGNO experiment

The previous analyses were conducted in dark conditions, without any light events, therefore, the measured performance does not account for any sensor characteristics related to light sensitivity, such as the influence of quantum efficiency on sensor performance. In this section, we use a prototype detector from the CYGNO experiment to generate well-defined light signals produced by particle interactions from a $^{55}Fe$ radioactive source, thus including performance measurements that account for such sensitivity.

## 4.1 $^{55}Fe$ radioactive source

To assess the impact of the sensors in the LIME detector, a $^{55}Fe$ radioactive source emitting 5.9 keV photons is employed, generating low-light signals representative of those expected during CYGNO operation. The energy of these photons is fully absorbed in the gas after traveling a few millimeters inside the detector, producing small spots in the generated image [12], as shown in Figure 15, where approximately a dozen such signals are present. The elongated tracks are due to ambient radiation, while the events generated by the $^{55}Fe$ source can be characterized as small circular spots [13].

To reconstruct the events, the CYGNO reconstruction algorithm [14] was used. This algorithm forms *clusters* in the image and extracts parameters such as energy, width, and length, among others. Using the energy parameter and knowing that the photon events have an energy of 5.9 $keV$, we can obtain the conversion factor from *ADC counts* to *electron-volts* of the detection system and use it to convert the noise measurements to *electron-volts*, thereby more completely assessing the impact each sensor may have on the CYGNO detection system, including the light sensitivity of each sensor.

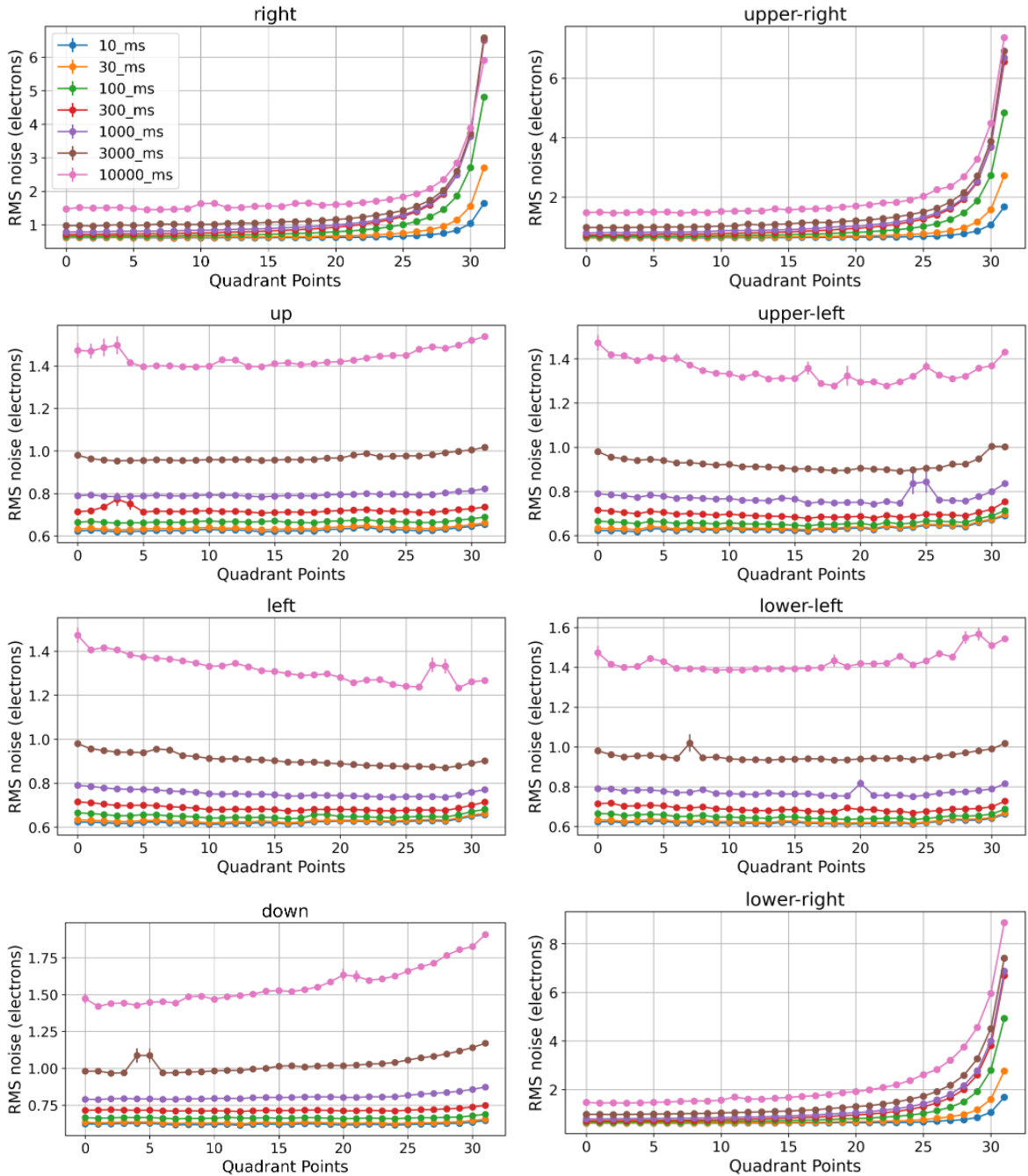

Figure 12: Variation of RMS noise across the eight angular regions for different exposure times of the Fusion-BT sensor.

Additionally, the obtained energy distributions can also be used to measure the energy resolution of the detection system when coupled with the different sensors under test. All measurements were performed under the same conditions. Only the sensors were exchanged so that the contribution of each sensor to the measured performance parameters could be isolated.

In the CYGNO reconstruction algorithm, each *cluster* is assigned an energy value calculated from the sum of the ADC counts of all pixels composing it. This intensity is expected to be proportional to the ionization produced by the particle in the gas and, therefore, to the released energy. The intensity distributions produced by interactions of the $^{55}Fe$ photons are shown in Figure 16 for the Fusion-BT and Quest sensors. These distributions can be modeled by two exponential functions, representing two components of background noise: one related to clusters formed by the sensor's electronic noise, concentrated in the low-energy region, and another related to clusters formed by ambient radiation, covering a larger energy range. Clusters formed by photons from the radioactive source can be modeled by a Gaussian function. Therefore, to fit the resulting distribution, a function composed of the sum of these three components was considered, as shown by the solid line in Figure 16.

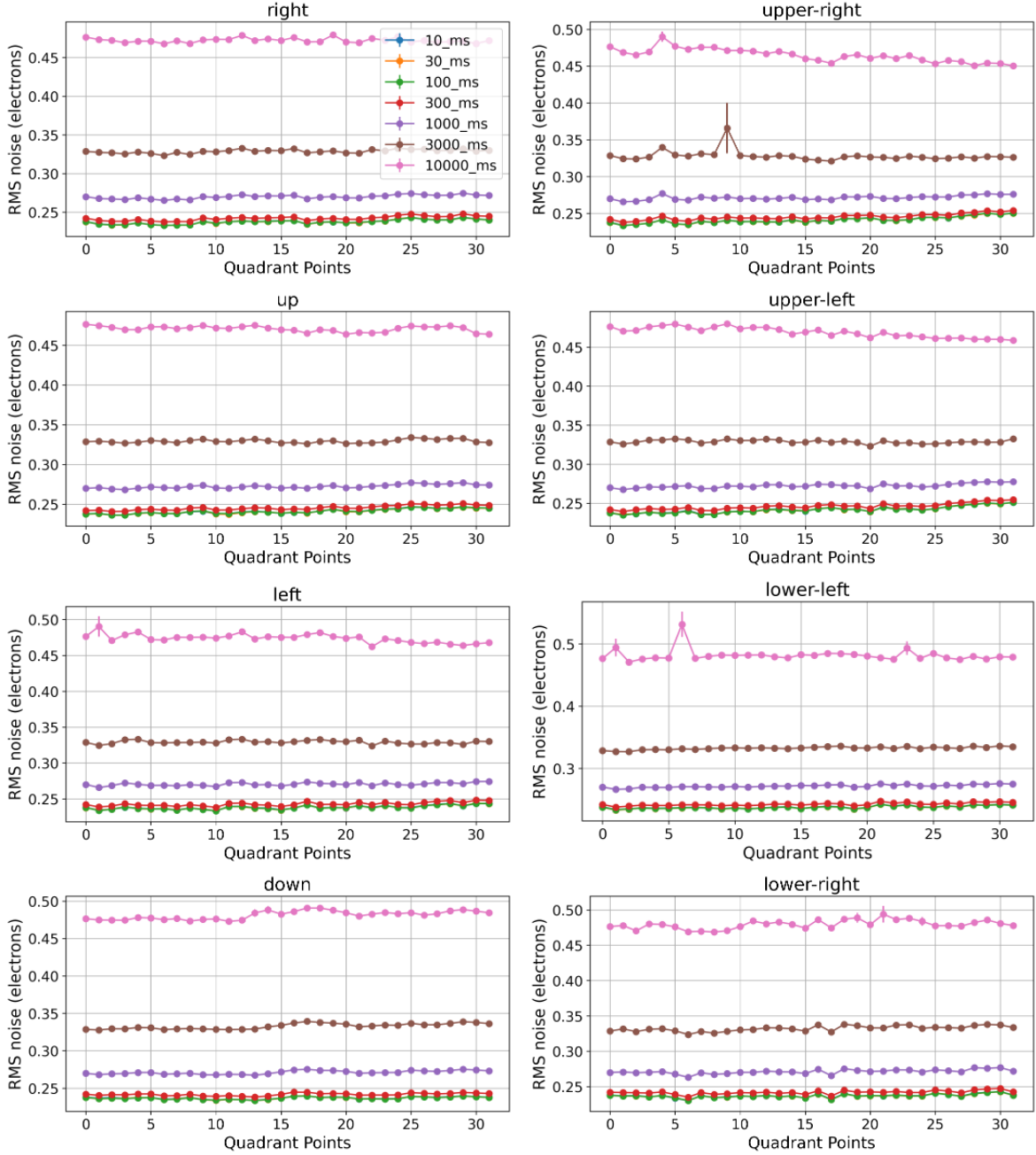

Figure 13: Variation of RMS noise across the eight angular regions for different exposure times of the Quest sensor.

Using these measurements, the conversion factors from *ADC counts* to *electron-volts* were obtained, being 0.53 and 0.65 *eV/ADC count* for the Fusion-BT and Quest, respectively. The measured energy resolutions were 13.0% and 10.1% for the Fusion-BT and Quest, respectively. It is also evident that for the same 5.9 $keV$ event, applying the sensors' *ADC-to-electron* conversion factor, the Fusion-BT produces on average 2667 *electrons*, while the Quest produces 976 *electrons*. Therefore, when operating with the Fusion-BT sensor, the system is 2.7 times more sensitive to events compared to operation with the Quest sensor. This result is in agreement with the expectation based on the two effects described in Sections 2.3 and 2.4: the fraction of photons received by each sensor (1.95) and the combined influence of quantum efficiency and *Plexiglass* transmission (1.31). When both effects are considered, they lead to an expected value of 2.6, which corresponds to the sensitivity ratio between the Fusion-BT and Quest sensors in terms of photoelectron production.

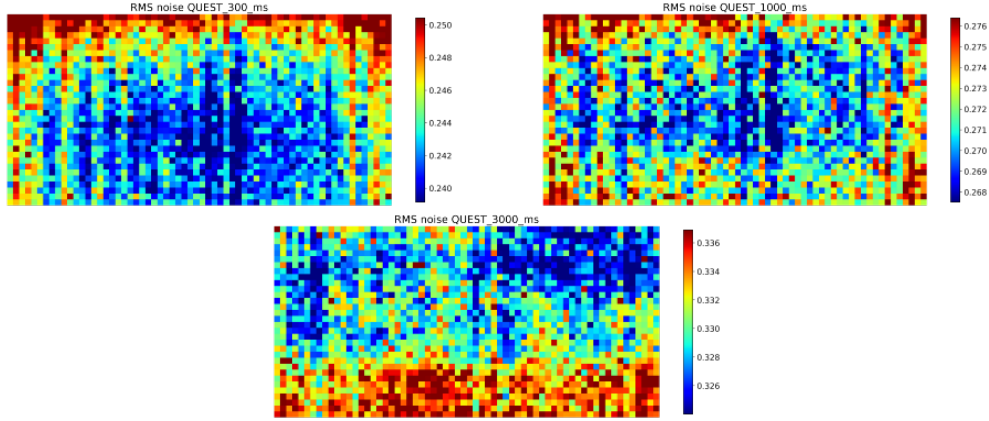

Figure 14: Variation of RMS noise across regions of the Quest active area for exposure times of 300 ms, 1000 ms and 3000 ms, respectively.

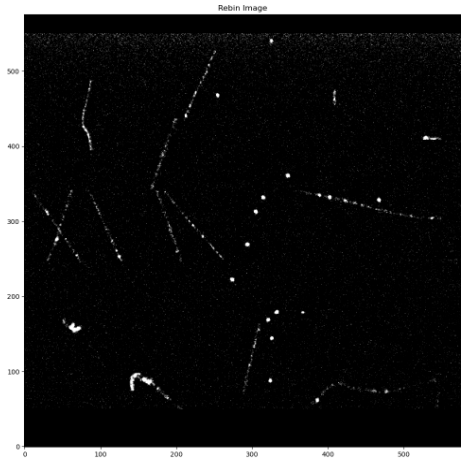

Figure 15: Image acquired in the presence of a $^{55}Fe$ radioactive source.

Consequently, although the Quest sensor has significantly better noise specifications, its distance from the GEM plane and its Quantum Efficiency make its performance in the CYGNO detection system—while still superior—closer to that of the Fusion-BT sensor, as will be demonstrated in the next section.

## 4.2 RMS noise measurements expressed in electron-volt units.

The RMS noise is reported in electron-volts (eV) to provide a direct link between sensor readout fluctuations and the energy deposited by particles in the gas. This representation allows a clearer evaluation of the detector's effective energy threshold and sensitivity.

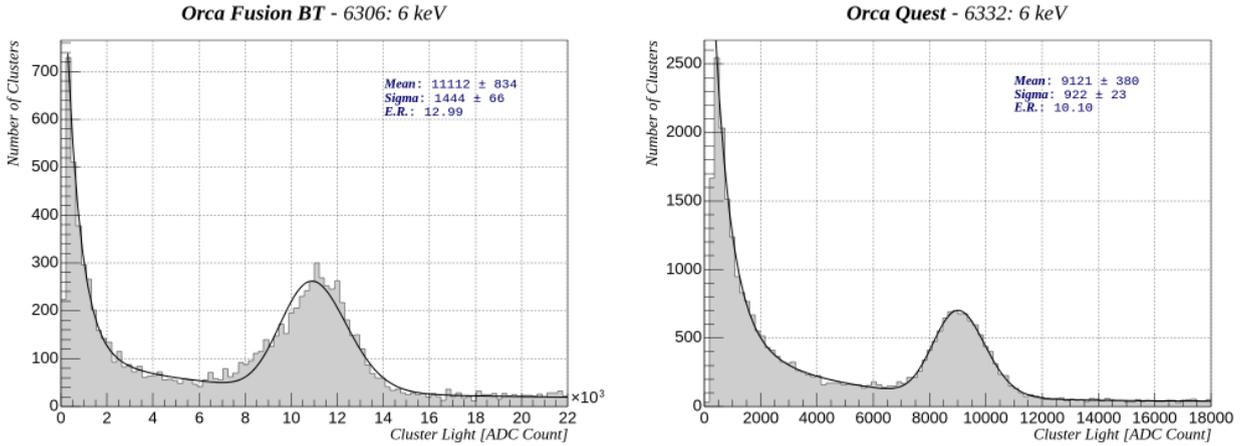

Figure 16: Energy distributions obtained with a $^{55}Fe$ radioactive source for the Fusion-BT and Quest sensors.

Figure 17 shows the RMS noise measurements converted to *electron-volts*. For a 10 *ms* exposure time, both sensors show similar RMS noise of 1.47 and 1.44 *eV* for the Fusion-BT and Quest sensors, respectively. As the exposure time increases, the Quest sensor exhibits progressively lower noise compared to the Fusion-BT, reaching a difference of 1.23 *eV*.

Thus, the improvement in Fusion-BT performance relative to Quest is evident when results are presented in *electron-volts*, compared to those presented in *electrons* (where the RMS noise of Quest is almost three times lower than that of Fusion-BT). This occurs as a result of the Fusion-BT sensor's considerably larger effective active area relative to the Quest, allowing it to be positioned closer to the GEM plane, in addition to its superiority in quantum efficiency. On the other hand, the Quest sensor, besides being superior in terms of *read noise*, has much lower *dark current* values, making it less affected by exposure time and resulting in better overall performance.

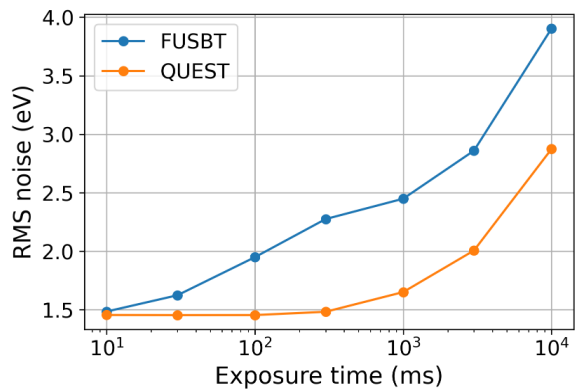

Figure 17: RMS noise in *eV* for the Fusion-BT and Quest sensors.

## 4.3 Comparison with the LIME nominal setup

One of the practical goals of this work is to identify sensors with the potential to enhance the performance of the CYGNO experiment's detection system. As previously mentioned, LIME, the latest prototype of the experiment, is nominally equipped with a Hamamatsu Fusion sensor.

With this configuration, LIME produced 1,990 electrons for 5.9 keV photon events [14], whereas the Fusion-BT sensor yielded 2,667 electrons under the same conditions. This difference is consistent with the distinct quantum efficiency characteristics of the two sensors (see Table 3).

The RMS noise of the Fusion sensor was evaluated for exposure times of 100 ms, 1,000 ms, and 10,000 ms. Tables 4 and 5 present the RMS noise values expressed in electrons and electron-volts, respectively, for the Fusion-BT, Quest, and Fusion sensors for these exposure times.

| Exposure Time | RMS Noise ($e^-$) | | |
|---|---|---|---|
| | Fusion-BT | Quest | Fusion |
| 100 ms | 0.88 | 0.24 | 0.85 |
| 1,000 ms | 1.11 | 0.27 | 1.14 |
| 10,000 ms | 1.76 | 0.43 | 2.39 |

The uncertainties are on the order of $10^{-4}$.

Table 4: RMS noise of the Fusion-BT, Quest, and Fusion sensors in *electrons* for different exposure times.

| Exposure Time | RMS Noise ($eV$) | | |
|---|---|---|---|
| | Fusion-BT | Quest | Fusion |
| 100 ms | 1.94 | 1.47 | 2.12 |
| 1,000 ms | 2.43 | 1.65 | 2.86 |
| 10,000 ms | 3.87 | 2.64 | 5.98 |

The uncertainties are on the order of $10^{-4}$.

Table 5: RMS noise of the Fusion-BT, Quest, and Fusion sensors in *electron-volts* for different exposure times.

Considering the measurements in *electrons*, the Quest is far superior to the other sensors, while the Fusion-BT performs similarly to the Fusion for exposure times up to 1000 $ms$. For 10000 $ms$, the Fusion shows a significant degradation compared to the Fusion-BT. In *electron-volts*, the Fusion performs significantly worse than the Quest, and compared to the Fusion-BT, its performance is slightly worse for 100 $ms$, deteriorating further with longer exposure times. This is justified by the lower quantum efficiency of the Fusion compared to the Fusion-BT, causing the gap between noise measurements to increase when converting from *electrons* to *electron-volts*. As previously stated, although positioned at a greater distance from the GEM plane, the Quest sensor demonstrated the best overall performance as a result of its very low noise characteristics.

## 4.4 CYGNO-04 considerations

The next phase of the CYGNO experiment focuses on constructing the CYGNO-04 demonstrator, designed to validate the feasibility of scaling to a final detector with a volume of order 30 L. The demonstrator consists of a back-to-back 0.4 $m^3$ TPC with a central cathode, operated with a $He/CF_4$ gas mixture at atmospheric pressure and room temperature. Signal amplification will be achieved using two Triple-GEM stacks, mounted on both readout planes, each having lateral and height dimensions of 500 mm and 800 mm, respectively. Each GEM surface will be optically readout by three Quest cameras, positioned 73 cm from the GEM plane, providing adequate optical coverage and light collection for the intended measurements.

Considering this distance from the GEM plane, the noise and sensitivity performance of the optical readout scheme can be estimated based on the measurements presented in the previous sections. At a distance of 73 cm from the GEM plane, CYGNO-04 is expected to produce approximately 42% more photoelectrons when compared to LIME when using the same Quest sensor and operated under comparable conditions, yielding an estimated average of about 1382 electrons for 5.9 keV photon events. Therefore, by converting the results obtained with the LIME detector to the CYGNO-04 configuration, the corresponding RMS noise values in electron-volts can be estimated, as reported in the last column of Table 6.

| Exposure Time | RMS Noise ($eV$) | | |
|---|---|---|---|
| | Fusion-BT | Quest$_{LIME}$ | Quest$_{CYGNO-04}$ |
| 10 ms | 1.484 | 1.455 | 1.023 |
| 30 ms | 1.623 | 1.454 | 1.022 |
| 100 ms | 1.949 | 1.454 | 1.022 |
| 300 ms | 2.275 | 1.482 | 1.042 |
| 1,000 ms | 2.449 | 1.651 | 1.161 |
| 3,000 ms | 2.859 | 2.006 | 1.410 |
| 10,000 ms | 3.903 | 2.873 | 2.020 |

The uncertainties are on the order of $10^{-4}$.

Table 6: RMS noise in electron-volts for Fusion-BT and Quest sensors with LIME, and an estimated value for Quest with CYGNO-04, for all considered exposure times.

Such values indicate that, in comparison with Fusion-BT, Quest when operated with CYGNO-04 exhibits a reduction in RMS noise ranging from approximately 30% at short exposure times to nearly 50% at long exposures. A consistent reduction of about 30% is likewise observed relative to Quest when operated with LIME, assuming identical operational conditions. These findings underscore the superior performance of the CYGNO-04 configuration in enhancing the overall sensitivity of the detection system.

# 5 Final Considerations

This work presents detailed measurements of two state-of-the-art scientific CMOS sensors—ORCA-Fusion-BT and ORCA-Quest—both developed by Hamamatsu, in the CYGNO experiment context, designed for rare, low-energy particle detection. The Fusion-BT sensor is characterized by its high quantum efficiency, while the Quest sensor stands out for its remarkably low noise performance. Two main aspects were investigated. The first concerns dark signal characterization, carried out through a comprehensive evaluation of pedestal and RMS noise properties across seven exposure times, ranging from 10 ms to 10 000 ms. Variations in distribution and spatial uniformity with increasing exposure time were examined in detail.

The RMS noise of the Fusion-BT sensor varied from 0.77 to 1.75 electrons, while the Quest ranged from 0.25 to 0.49 electrons. The Fusion-BT showed an immediate increase in RMS noise with exposure time, whereas the Quest maintained a nearly constant value of 0.25 electrons up to 300 ms, rising slightly to 0.28 electrons at 1 000 ms. In terms of spatial behavior, the Fusion-BT exhibited pronounced non-uniformity along the right border of the active area, visible in both pedestal and RMS noise maps. In contrast, the Quest demonstrated excellent uniformity—within 5% variation—except for a small region at the top edge, where pedestal values slightly increased.

The second aspect examined the performance of these sensors when integrated into an optical-readout GEM-based detector operating in low-light conditions. In this context, a $^{55}Fe$ radioactive source emitting 5.9 keV photons was employed to incorporate the light sensitivity of each sensor into the analysis. This was achieved by converting the RMS noise units from electrons to electron-volts, a representation that accounts for both intrinsic electronic noise and the photon-detection efficiency of each device. When integrated into the LIME prototype, the detector achieved an energy resolution of 13.0% with the Fusion-BT sensor and 10.1% with the Quest sensor. Measurements based on the $^{55}Fe$ source showed that, while the Quest's lower intrinsic RMS noise provides a clear advantage, this benefit is partially offset by the Fusion-BT's higher quantum efficiency and larger active area, which allow it to be positioned closer to the light-generating GEM plane. For instance, at 10 ms, both sensors exhibited comparable RMS noise levels of approximately 1.5 eV. However, since the Quest sensor is less sensitive to increases in exposure time, it yields better RMS noise performance than the Fusion-BT for longer exposures. As an example, for 300 ms, the Quest maintained 1.5 eV of RMS noise compared with 2.3 eV for the Fusion-BT.

For completeness, a comparison was also made with the previous-generation ORCA-Fusion sensor, currently used in CYGNO for system validation. Both newer models, Fusion-BT and Quest, demonstrated clear performance improvements over the older Fusion device, mainly for higher exposure times. Finally, an estimate of the RMS noise in electron-volts units expected for the next CYGNO detector under construction, CYGNO-04, was provided. When equipped with the Quest sensor and positioned closer to the GEM plane, as proposed for CYGNO-4, the configuration is expected to yield 42% more photoelectrons than that of LIME, as a consequence of the reduced sensor distance, resulting in a 30%-50% reduction in RMS noise.